\documentclass[final,5p,times]{elsarticle} 
\usepackage{amssymb} 
\usepackage{amsthm}
\usepackage{amsmath}
\usepackage{hyperref,graphicx}\hypersetup{colorlinks = true}
\usepackage{lineno}
\modulolinenumbers[5]
\DeclareGraphicsExtensions{.eps,.png,.jpg,.pdf}
\journal{Elsevier}

\begin{document}
\begin{frontmatter}
\title{Multiple-Photon Disambiguation on Stripline-Anode Micro-Channel Plates}

\author[ultralytics]{Glenn R. Jocher}\corref{cor1}\ead{glenn.jocher@ultralytics.com}
\author[efi]{Matthew J. Wetstein}\ead{mwetstein@uchicago.edu}
\author[Incom]{Bernhard Adams}\ead{badams@incomusa.com}
\author[ultralytics]{Kurtis Nishimura}\ead{kurtis.nishimura@ultralytics.com}
\author[ngaLong,GMU]{Shawn M. Usman}\ead{shawn.usman@nga.mil}

\address[ultralytics]{Ultralytics LLC, Arlington, VA 22203, USA}
\address[efi]{Enrico Fermi Institute, University of Chicago, 5640 S. Ellis Ave., Chicago, IL 60637, USA}
\address[ngaLong]{Research Directorate, National Geospatial-Intelligence Agency, 7500 GEOINT Dr., Springfield, VA, 22150,~USA}
\address[GMU]{Department of Geography and Geoinformation Science, George Mason University, Fairfax, VA 22030,~USA}
\address[Incom]{Incom, Inc., 294 Southbridge Road, Charlton, MA 01507, USA}
\cortext[cor1]{Corresponding author}

\begin{abstract}
Large-Area Picosecond Photo-Detectors (LAPPDs) show great potential for expanding the performance envelope of Micro-Channel Plates (MCPs) to areas of up to 20$\times$20cm and larger. Such scaling introduces new challenges, including how to meet the electronics readout burden of ever larger area MCPs. One solution is to replace the traditional grid anode used for readout with a microwave stripline anode, thus allowing the channel count to scale with MCP width rather than area.

However, stripline anodes introduce new issues not commonly dealt with in grid-anodes, especially as their length increases. One of these issues is the near simultaneous arrival of multiple photons on the detector, creating possible confusion about how to reconstruct their arrival times and positions. We propose a maximum {\it a posteriori} solution to the problem and verify its performance in simulated scintillator and water-Cherenkov detectors.\end{abstract}

\begin{keyword}
Stripline \sep Anode \sep  Micro-Channel Plate (MCP) \sep Large Area Picosecond Photo-Detector (LAPPD) \sep  Photo-electron (PE)
\end{keyword}

\end{frontmatter}
\tableofcontents

\section{Introduction}

In an LAPPD device, photoelectrons from a photocathode are amplified in a stack of Micro-channel Plates (MCP), and the resulting charge cloud is deposited on a stripline-array anode. On each of the affected striplines, the electric charge propagates from the origination point in two opposite directions to the ends of the striplines where the pulse waveforms are acquired with fast analog-to-digital converters (ADCs).

\begin{figure}[!htbp]\centering
\includegraphics[width=1\linewidth]{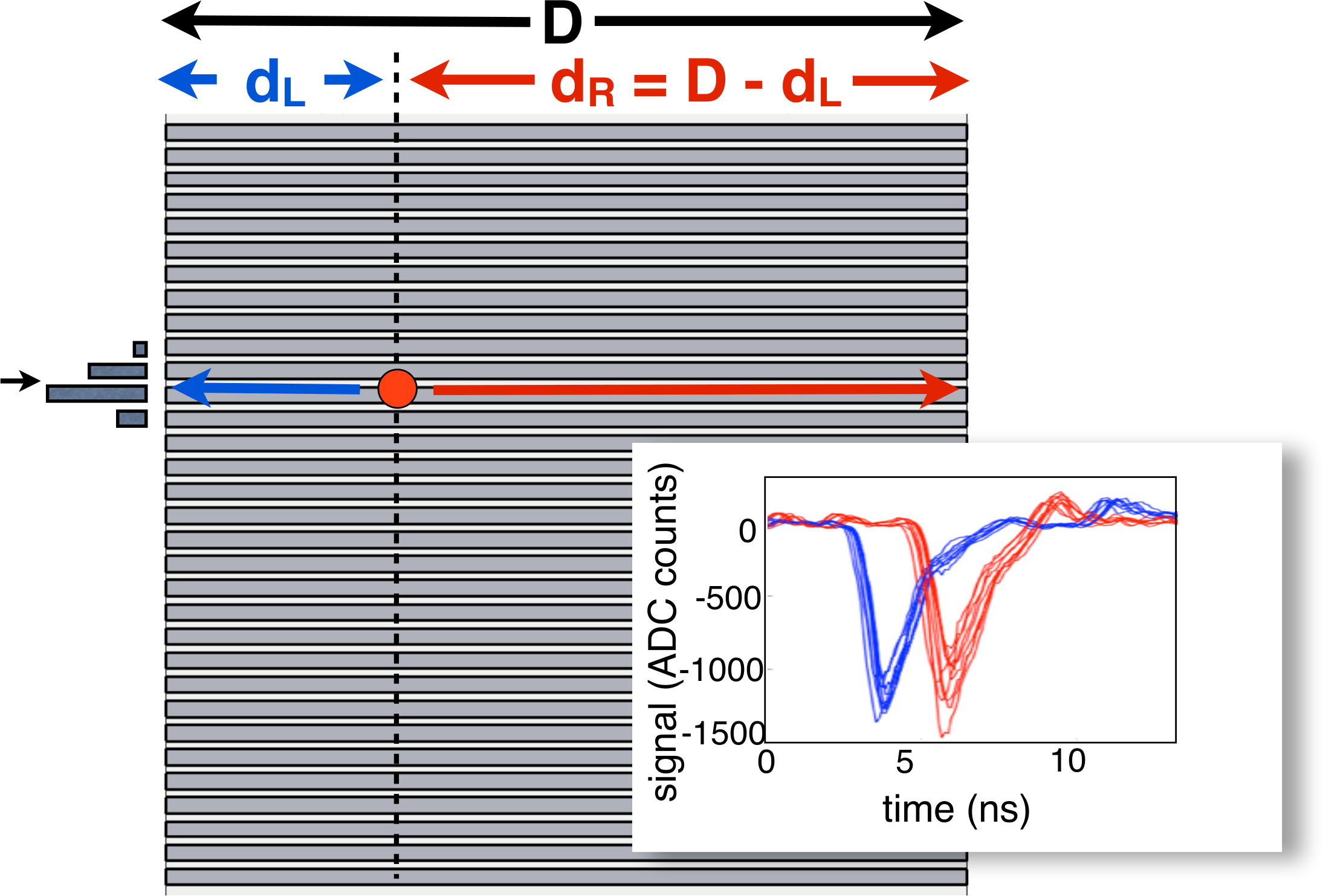}
\caption{The stripline-anode pattern on an LAPPD, consisting of 30 4.62 mm-wide striplines separated by 2.29 mm gaps. PE locations are determined by the difference in arrival time between the two ends of the strips. Transverse position is determined by a charge centroid.}
\label{lappdexample}
\end{figure}

\begin{figure*}[!htbp]\centering
\includegraphics[width=.7\linewidth]{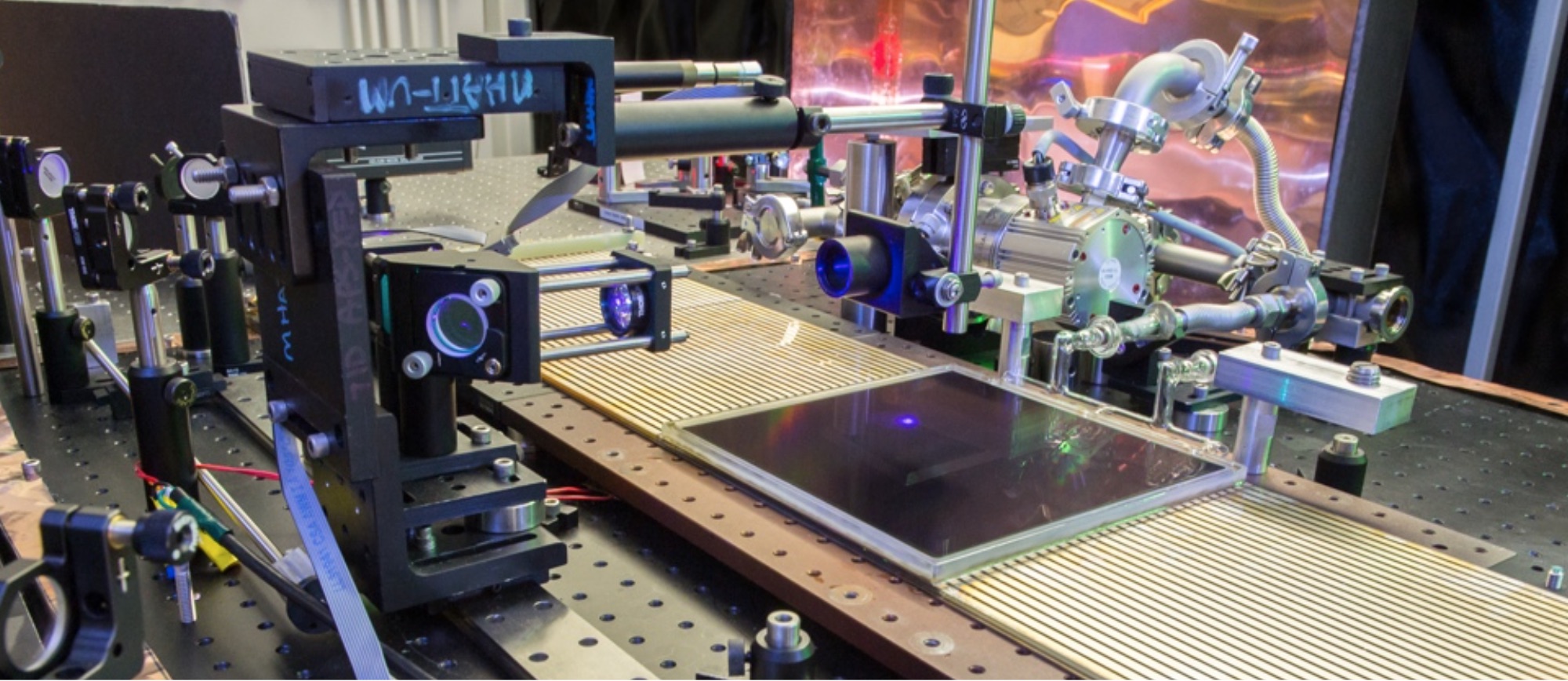}
\caption{LAPPD + PSEC4\cite{PSEC4 Paper} test setup at Argonne National Lab\cite{test setup}.}
\label{labsetup}
\end{figure*}

The timing of the original charge-deposition event is found as the average of the two pulse-arrival times, and the position along the striplines is found by their difference, scaled with the signal propagation speed $v$ (typically 0.6 $c$, where $c$ is the vacuum speed of light, depending on the dielectric properties of the substrate). In the other direction, the position of the detection event is found by interpolating signal strengths between striplines. Stripline anodes and associated readout strategies have been discussed in the past works such as Lampton {\it et al}\cite{Lampton_1987} and Jagutzki {\it et al}\cite{Jagutzki_2007}, with early developments tracing back to work on proportional chambers (e.g., Rindi {\it et al}\cite{Rindi_1970} and Grove {\it et al}\cite{Grove_1970}). The anode geometry studied in this paper was developed specifically for the LAPPD project~\cite{anodepaper} and consists of 30 active striplines. In possible modifications to the original design, commercial LAPPD$^{\rm TM}$ detectors might make use of the outermost striplines for high voltage control. The following work therefore assumes only 26 active delay lines, albeit with the same pitch and spacing. \\


The event time $t$ and longitudinal position $d$ of the arrival PE can be determined from the following set of equations:
\begin{eqnarray}
\label{eq2_1}
t &=& \frac{t_L+t_R - D/v}{2}\\
\label{eq2_2}
d_L &=& D/2 - v\left(\frac{t_L-t_R}{2}\right)\\
\label{eq2_3}
d_R &=& D - d_L
\end{eqnarray}


\noindent where $t_L$ and $t_R$ are the left and right side arrival times of the charge cloud and D is the total strip length (20 cm in our LAPPD example). While this method works well for resolving single PEs, multiple pulses due to near-simultaneous PEs lead to multiple pulses arriving at the two sides of a stripline, and ambiguities about time and location may result.


For $n$ left-side pulses and $m$ right-side pulses, we create a likelihood matrix $\Lambda$ to evaluate all possible left-right pair combinations $\Lambda _{ij}$, where $i=1 \hdots n$  and $j=1 \hdots m$:

\begin{eqnarray}
\label{eqn3}
\Lambda = \left[ \begin{array}{ccc}\Lambda_{11} & \hdots & \Lambda_{1m}	\\
\vdots & \ddots & \vdots		\\
\Lambda_{n1} & \hdots & \Lambda_{nm} \end{array} \right]
\end{eqnarray}

Individual pair likelihoods $\Lambda_{ij}$ are the product of 3 component likelihoods: amplitude likelihood $\Lambda_a$, time probability $P_t$, and location likelihood $\Lambda_y$:
\begin{equation}
\Lambda_{ij} = \Lambda_a\left(a_L^i\mid a_R^j\right) P_t\left(t_L^i\mid t_R^j\right) \Lambda_y\left(y_L^i\mid y_R^j\right)
\end{equation}

The highest likelihood elements in each row (if there are less rows than columns) or column (if there are less columns than rows) are used to match up each left-side pulse with the most similar right side pulse. Pair likelihoods must exceed a certain threshold to reduce false positives. This threshold is determined {\it a priori} by Monte Carlo modeling of the LAPPD in it's intended target environment.

\section{Our Test Setup}

We assembled a laser test\cite{test setup} at the Argonne National Lab (ANL) Advanced Photon Source (APS) to gather empirical data on LAPPD performance. The setup, pictured in Figure \ref{labsetup}, consists of a prototype LAPPD with a high gain photo-cathode attached to continuously operating vacuum pumps. The prototype LAPPD had two MCP layers with 20 $\mu$m pores at an 8$^\circ$ bias angle, with 60 L/D ratio (1.2mm MCP thickness) and a 2.29mm gap between the second MCP and the anode\cite{test setup}. The LAPPD voltages are set by the relative resistances in each stack:

\begin{itemize}  \itemsep1pt \parskip0pt \parsep0pt
  	\item 5 M$\Omega$ across the photocathode gap
  	\item 5 M$\Omega$ across the gap between the MCPs
	\item 10 M$\Omega$ across the anode gap
	\item 40 M$\Omega$ across each MCP
\end{itemize}

At 2700 V High Voltage (HV), that means there would be 135 V across the photocathode gap and inter-MCP gap and 270 V across the anode gap.

A pulsed laser delivered photons to a targeted location on the LAPPD surface, and an external trigger fed 10 PSEC4\cite{PSEC4 Paper} chips, 5 on each side, which each recorded 256 samples along 6 channels of 12 bit digitized (10.5 Effective Number of Bits, (ENOB)\cite{PSEC4 Paper}) data covering all 26 anode-strips (both sides) on the LAPPD. The sampling rate was 10 Gs/s, or 100 ps between samples, for a total listening time of 25.6 ns. The noise level observed on the PSEC4 was about 1 mV with saturation occurring at +/- 1.1 V, equivalent to 3 PEs. Each laser pulse produced 2$\times$26 waveforms, each with 256 samples digitized at 10 bits.

We used multiple laser pulses to characterize the LAPPD Single Photon Response (SPR), including the distribution of the charge cloud signals (both along and across the anodes), amplitude distribution,  propagation speed $v$ down the strips ($v\approx 0.6 c$), and 20 $\mu$m MCP Transit Time Spread (TTS) of $\sim$60 ps\cite{LAPPDtiming} 1$\sigma$. We then modeled MonteCarlo (MC) SPRs from these distributions, and assumed analog linearity when extrapolating to a multiple photon response (single-PE charge clouds with average amplitudes of about 400 mV were linearly added if they overlapped, though our PSEC4 digitizer model ensured the total signal was clipped at +/-1.1 V, or about 3 PEs.). Our validated SPR is employed throughout this paper to model both single and multiple photons on an LAPPD.

\begin{figure}[!htbp]\centering
\includegraphics[width=1\linewidth]{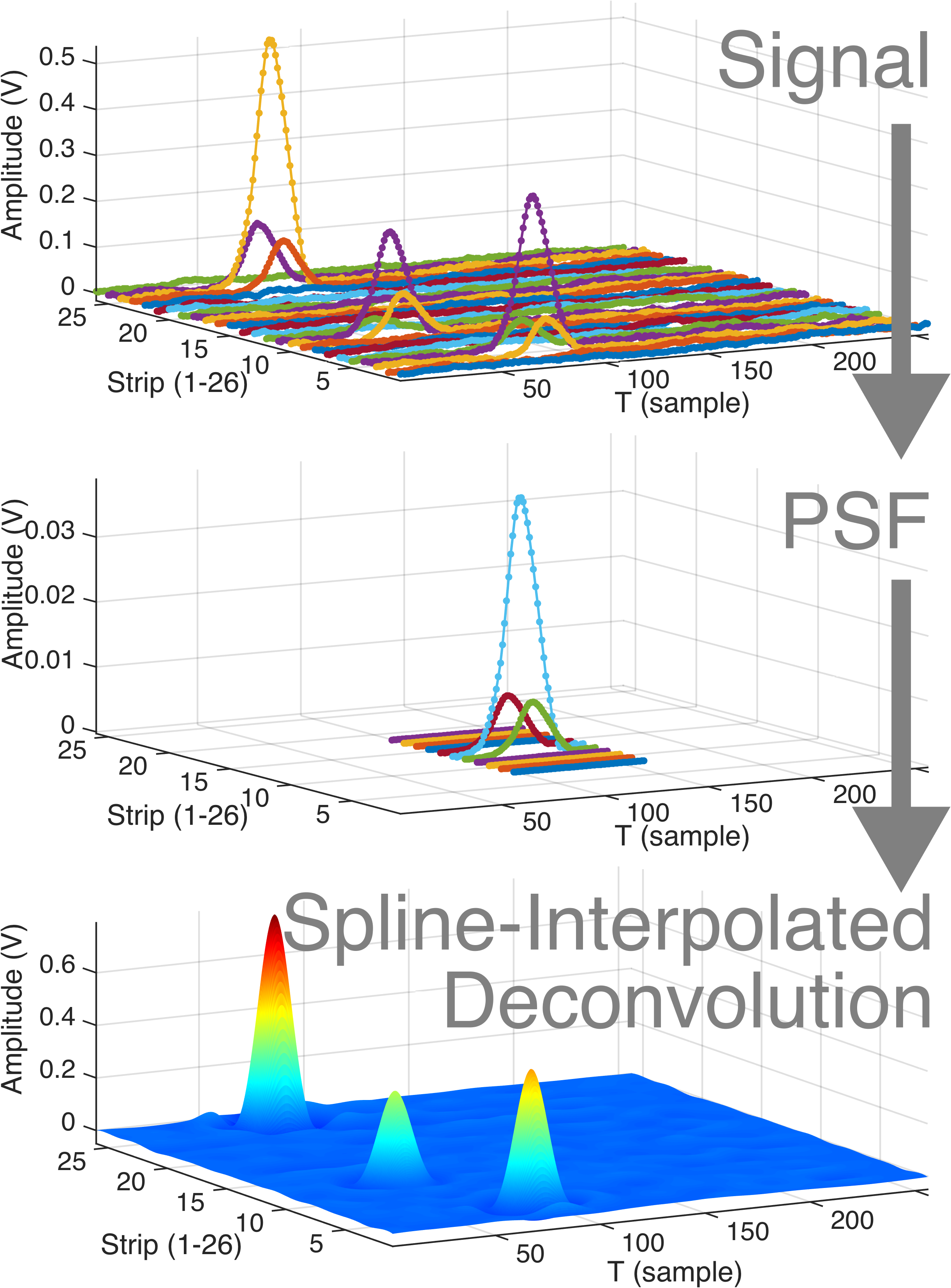}
\caption{{\bf UPPER:} Example ADC Signal on one side of the LAPPD showing 3 PEs. 256 time samples are recorded for each of the 26 strips. {\bf MIDDLE:} Empirical SPR 2D Point Spread Function (PSF) used to deconvolve the Signal. $\sum_{\mathrm{PSF}}=1$. {\bf LOWER:} Signal deconvolved by PSF, followed by a fine spline interpolation over the entire region.}
\label{psf}
\end{figure}

\section{Signal Deconvolution and Interpolation}

Before matching left and right side pulses one must first estimate amplitudes $a^i_L$ and $a^j_R$, arrival times $t^i_L$ and $t^j_R$, and across-strip locations $y^i_L$ and $y^j_R$ of all left ($i=1...n$) and right ($j=1...m$) side pulses.

The first part of this process is to perform a 2-dimensional deconvolution of the signal by the LAPPD Point Spread Function (PSF). The PSF shape we use corresponds directly to the average charge-cloud shape deposited on the anode surface as measured in the lab. The motive behind the deconvolution is to help separate overlapping pulses. Our simulation results show that single-PE performance remains unaffected by this deconvolution, while deconvolution of higher occupancy events generally provides for greater disambiguation efficiency and PE resolution when compared to non-deconvolved results.

For our LAPPD geometry our signal is mapped onto 256 time samples per each of 26 striplines, which we represent as a 26$\times$256 signal matrix $S$ on each side of the LAPPD. To deconvolve this noisy data we use a 2D Wiener\cite{WNR} deconvolution. A typical Noise to Signal Ratio (NSR) in our example application is about $(4 mV)^2/Var(S)$, assuming a signal noise of 4 mV $1\sigma$ and a total signal $S$ variance of $Var(S)$.

An example signal matrix, its associated PSF, and the resulting deconvolution using this PSF are shown in Figure \ref{psf}.  The PSF is created from empirical measurements of charge cloud distributions from the APS test setup.  The signal model $S$ is constructed as the sum of 3 MonteCarlo PEs.  This matrix is deconvolved and spline interpolated to resample the data by a factor of 10 in voltage and in time.  We use the resulting 260$\times$2560 data points to resolve local maxima with finer precision than the original deconvolved signal.

\section{Disambiguation Methods}

\subsection{Pulse Time}

Equation \ref{eqnPt}, a Normal Cumulative Distribution Function (CDF), defines the probability (from 0 to 1) that two pulses {\bf may} have originated from the same source based solely on left-right pulse arrival times $t_L$ and $t_R$, where $D$ is the stripline length (20cm in this example), $v$ is the pulse speed down the stripline (v$\approx 0.6 c$), and $\sigma = 15$ ps is the assumed measurement resolution of each $t$.
\begin{equation}
P_t = 1 - \frac{1}{2}\left[1 + erf\left(\frac{\left|t_L-t_R\right| - \frac{D}{v}}{\sigma\sqrt{2}}\right) \right]
\label{eqnPt}
\end{equation}

\begin{figure}[!htbp]\centering
\includegraphics[width=1\linewidth]{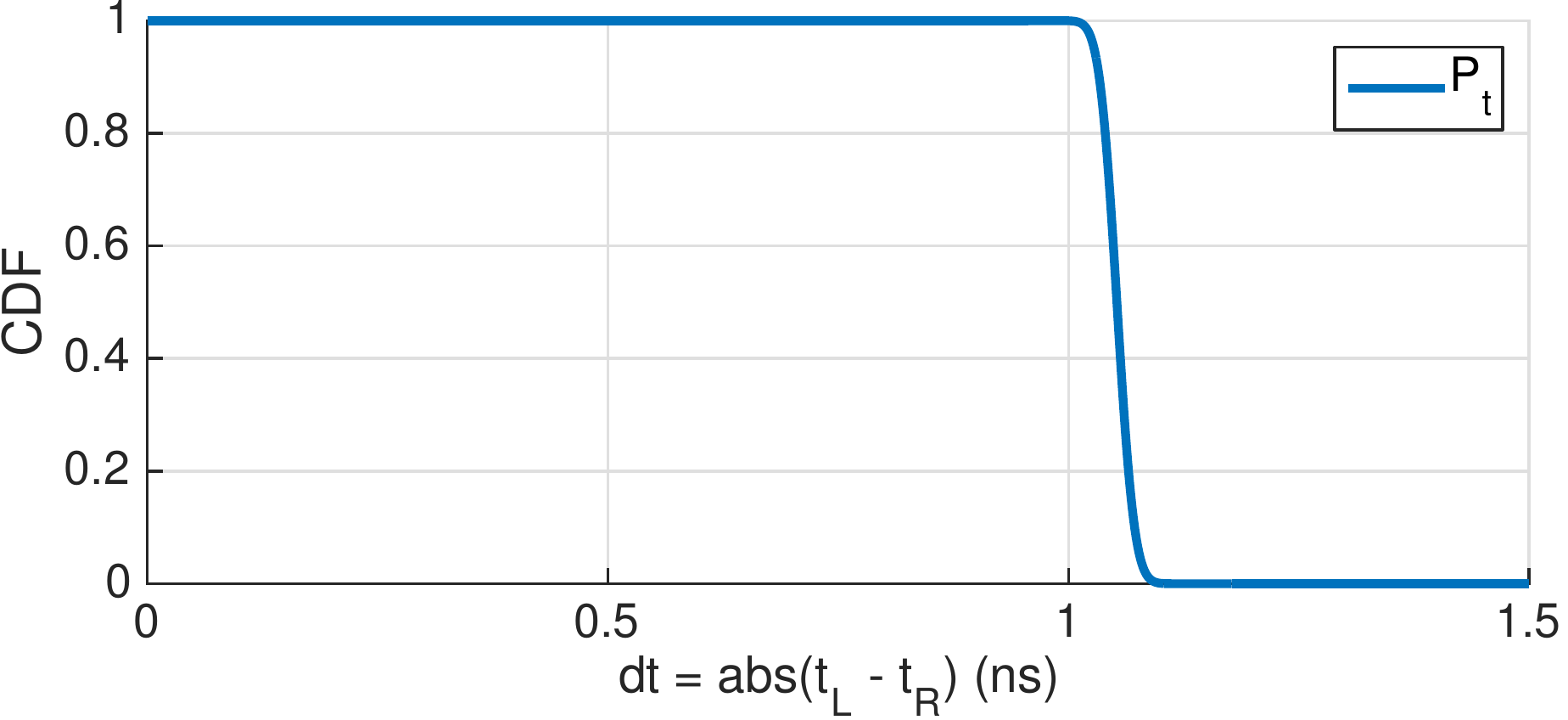}
\caption{Gap in time on the LAPPD between which two pulses may share a common source. $dt$ values beyond $\pm$1ns indicate low to zero probability of a match.}
\label{timegap}
\end{figure}

For the LAPPD we find that left-right pulses may only share a common source if $t_L$ and $t_R$ fall within about 1 ns of each other. This finding is based on an experimentally measured pulse propagation speed of about $v\approx 0.6 c \approx 180$ mm/ns, and an anode-strip width of 200 mm. Longer time intervals indicate low likelihood the pair could have been created by the same PE. The $\sigma = 15$ ps smear in Equation \ref{eqnPt} accounts for errors introduced when measuring $t_L$ and $t_R$, and will vary based on the exact timing characteristics of the MCP being used.

\subsection{Pulse Amplitude}

\begin{figure}[!htbp]\centering
\includegraphics[width=1\linewidth]{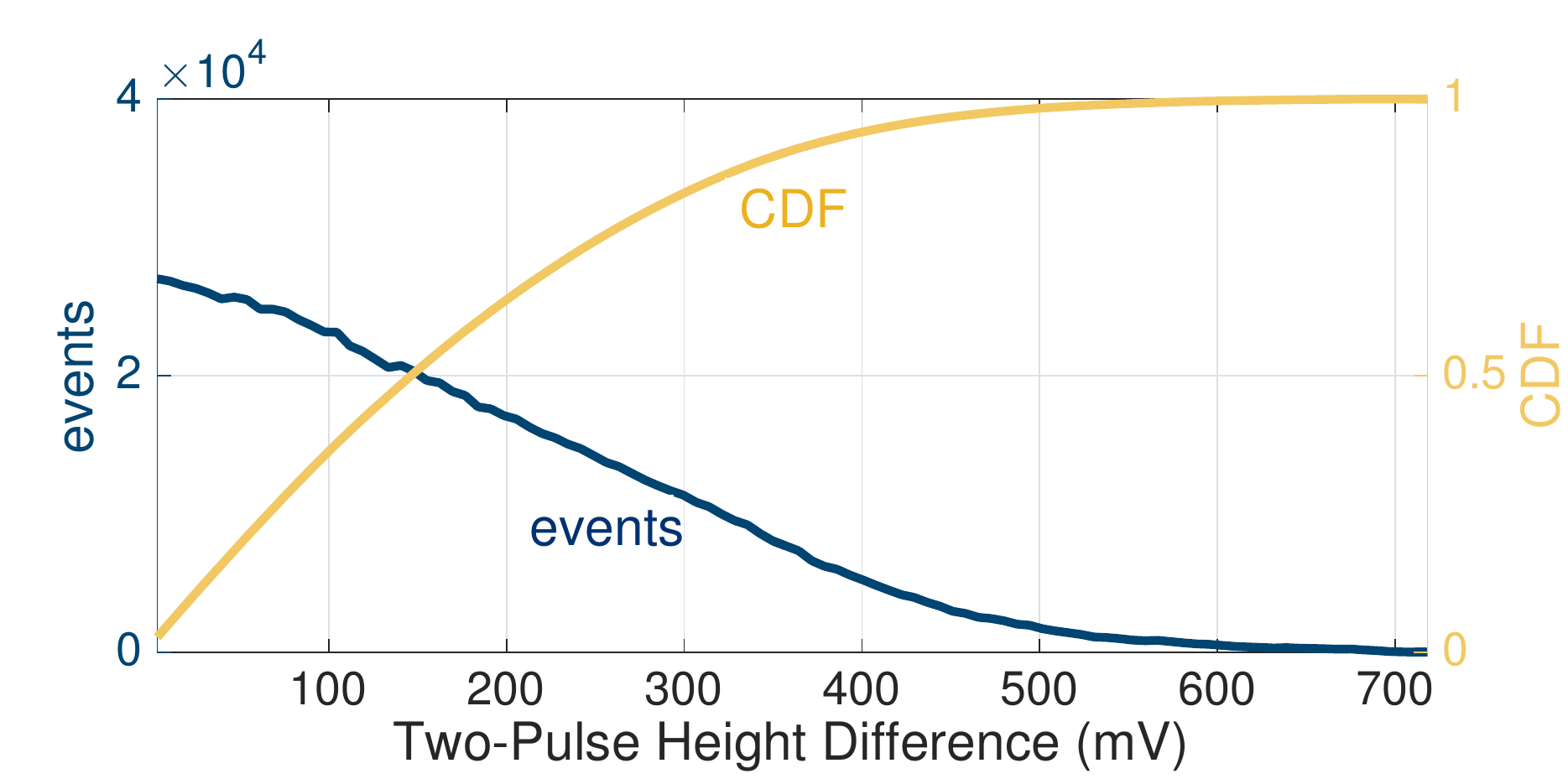}
\caption{Measured amplitude differences between random pairs of single-PE pulses on the LAPPD, indicating a low probability of encountering two random pulses of similar amplitude. Results shown are in anticipation of low rate applications where MCP saturation effects are not expected to be significant}
\label{ampdiff}
\end{figure}

We use Equation \ref{eqnPa}, a Normal Probability Density Function (PDF), to define the likelihood that two pulses originated from the same source based solely on the observed left and right pulse amplitudes $a_L$ and $a_R$. Measured pulse amplitude differences are shown in Figure \ref{ampdiff} for random pairs of single PEs on the LAPPD.

$\sigma$ in Equation \ref{eqnPa} represents the left-right amplitude-difference resolution for two pulses on the same strip (created by the same PE). This makes $\sigma$ a {\it tunable parameter}, tunable to the resolutions encountered at different occupancies (higher occupancies incur charge cloud overlap which worsens this two-pulse amplitude difference resolution). Higher $\sigma$ values will allow matching of left-right pulses with larger amplitude differences, thus higher $\sigma$ values will be more suited to high occupancy environments where charge cloud overlap is more common, while lower $\sigma$ values are better suited to low occupancy environments where charge cloud overlap is more rare. The value of $\sigma$ is constrained at the lower end by two single PE amplitude resolutions (the best possible) added in quadrature. In this paper we assume $\sigma=75$ mV, a compromise designed to help deconvolve occupancy levels of about 1 PE per LAPPD strip anode.

\begin{equation}
\Lambda_a = \frac{1}{\sigma \sqrt{2 \pi}} e ^{\frac{\left(a_L-a_R\right)^2}{2 \sigma^2}}
\label{eqnPa}
\end{equation}

The pulse amplitudes themselves are measured as the local maximum of the Figure \ref{psf} spline. The threshold we use in declaring a local maxima in this example is $\geq$10\% of the maximum signal value, or about 100 mV. Values below this are ignored as they may be caused by edge ringing from the Wiener deconvolution, visible in the lower plot of Figure \ref{psf}.

\subsection{Pulse Location}

Equation \ref{eqnPy}, a Normal PDF, defines the likelihood that two pulses originated from the same source based solely on the locations of the measured pulse centroids $y_L$ and $y_R$ across the strips:

\begin{equation}
\Lambda_y = \frac{1}{\sigma \sqrt{2 \pi}} e ^{\frac{\left(y_L-y_R\right)^2}{2 \sigma^2}}
\label{eqnPy}
\end{equation}

These centroids are measured as the peaks of the spline matrix pulses. As the distance between $y_L$ and $y_R$ increases, the likelihood that they share a common source decreases. $\sigma$ represents the assumed measurement resolution of the across-strip centroid solutions $y_L$ and $y_R$. In this example we use $\sigma=4$ mm,  determined from empirical results.

\section{Modeled Scenarios}
\subsection{Neutron Scintillation Detector}

\begin{figure}[!htbp]\centering
\includegraphics[width=1\linewidth]{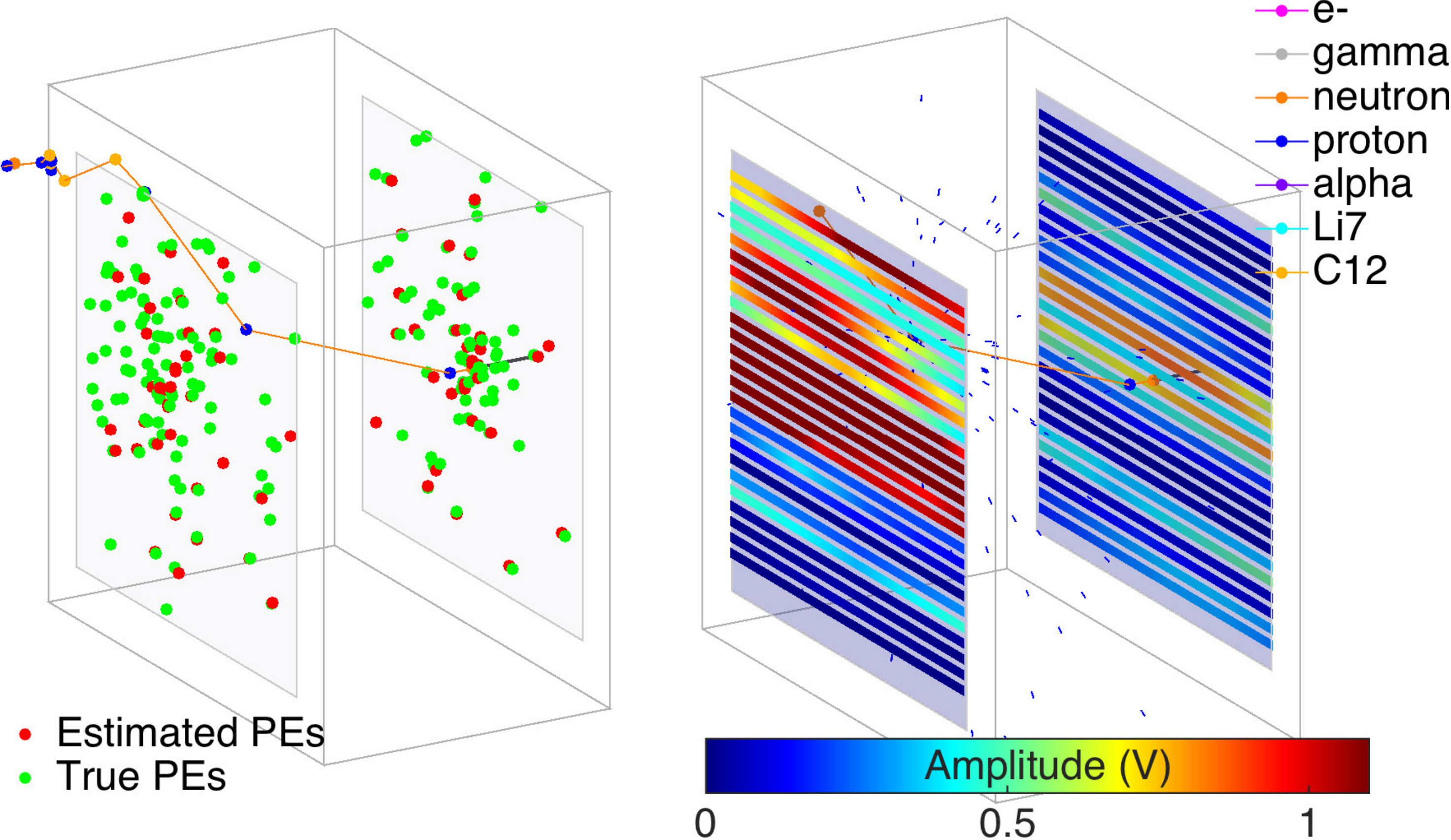}
\centering
\caption{Simulated event display of a 3 MeV neutron in a 25$\times$25$\times$15 cm 2-LAPPD EJ-25 plastic scintillator neutron detector. {\bf RIGHT:} Image captured 5 ns after the neutron enters the detector. About 200 PEs are observed on both LAPPDs combined. {\bf LEFT:} Example PE reconstructions (red dots). About half of the deposited PEs (green dots) are successfully disambiguated in this example.}
\label{NTC}
\end{figure}

\begin{figure}[!htbp]\centering
\includegraphics[width=1\linewidth]{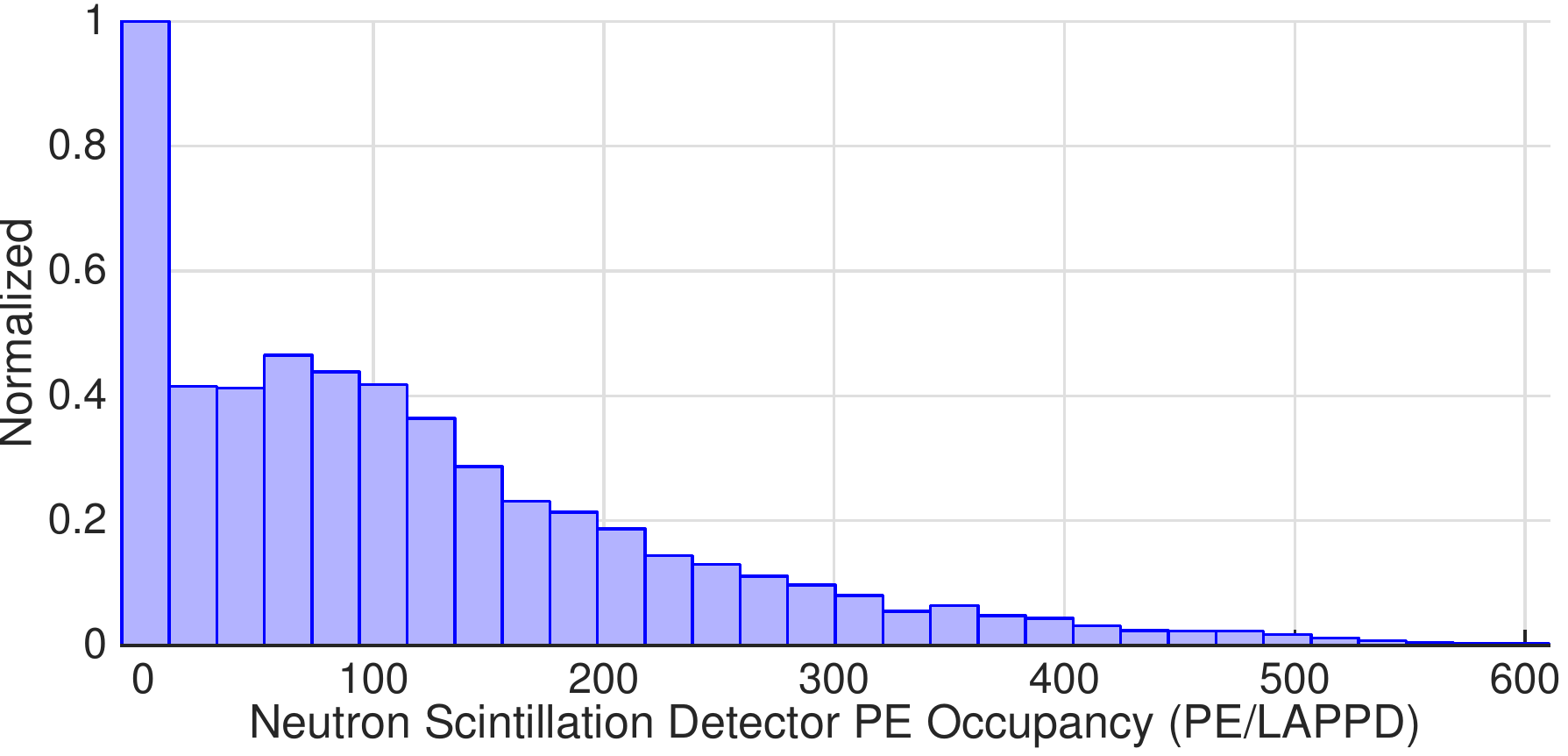}
\caption{GEANT4\cite{GEANT} and MATLAB\cite{MATLAB} MonteCarlo results of LAPPD PE occupancy for 3 MeV neutrons in the neutron scintillation detector, showing that most LAPPDs receive about 100-200 PEs per event.}
\label{NTC_occupancy}
\end{figure}

Our neutron scintillation detector example, shown in Figure \ref{NTC}, is essentially a small $\sim$MeV neutron scatter camera made from a single piece of Eljen EJ-254\cite{EJ-254} plastic scintillator sandwiched between 2 LAPPDs. The detector is 25$\times$25$\times$15 cm. We model particle kinematics in GEANT4\cite{GEANT} and photon, MCP and digitizer responses in MATLAB\cite{MATLAB}.

\begin{figure}[!htbp]\centering
\includegraphics[width=1\linewidth]{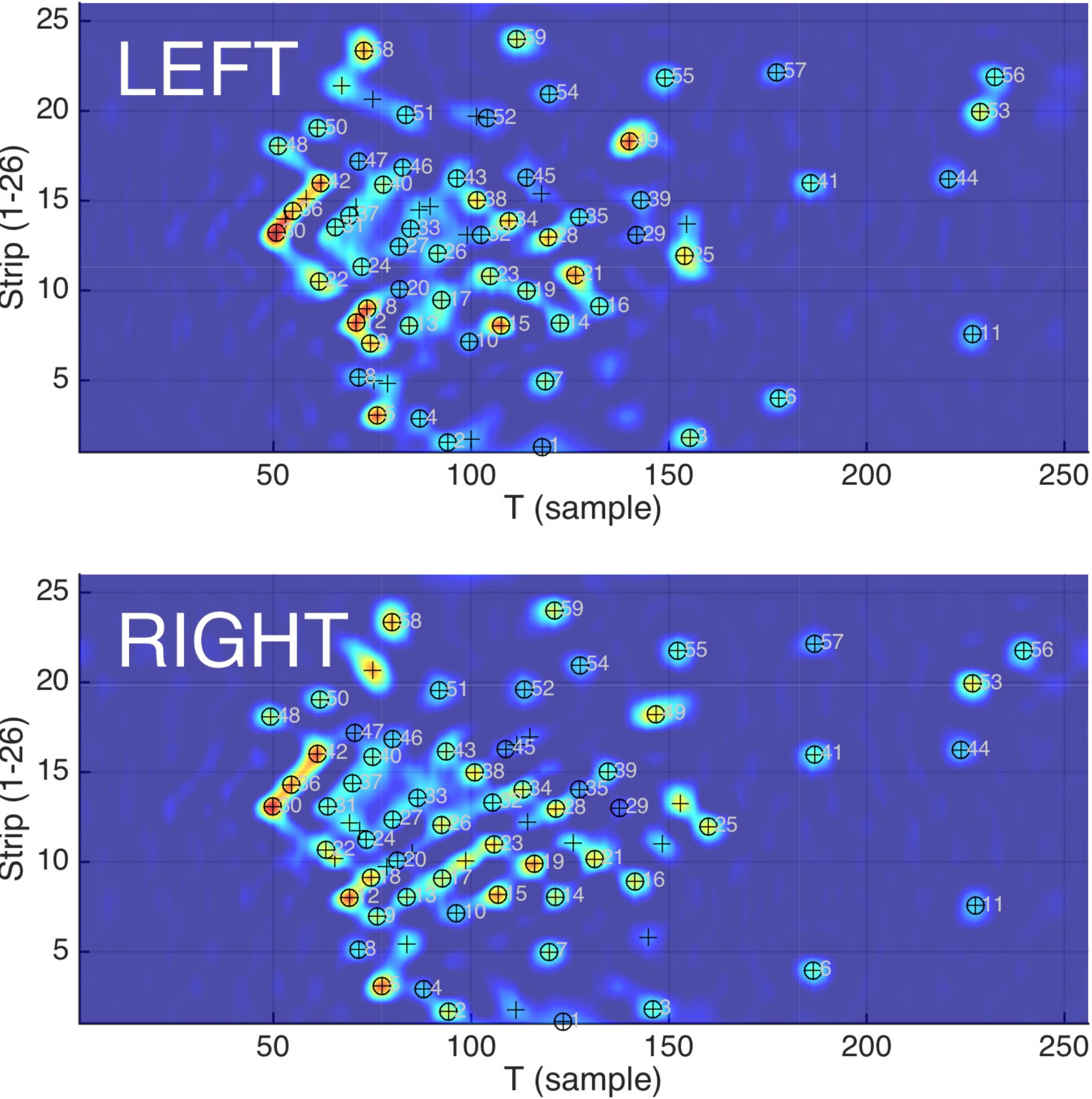}
\caption{Neutron scintillation detector disambiguation example showing 3 MeV neutron PE depositions on 1 single LAPPD (LEFT and RIGHT anode strip sides shown). Crosses indicate resolved PEs, and circles indicate successful left-right pairing of resolved PEs. 59 matches are seen here from the original 73$\times$75 pulse matrix $\Lambda_{ij}$. Match position and time resolutions are 14 mm and 80 ps at 42\% efficiency (59 matches / 141 PEs).}
\label{deconv3}
\end{figure}

In our example 3 MeV neutron event we see about 200 PEs on both LAPPDs. We examine one LAPPD on which 141 PEs are deposited. The left and right spline signal matrices $S_L$ and $S_R$ for the Figure \ref{NTC} example event are shown in Figure \ref{deconv3}.

Our disambiguation method located 73 left-side and 75 right-side spline maxima, and the amplitudes $a$, times $t$, and locations $y$ of all these maxima are passed onto a full 73$\times$75 $\Lambda_{ij}$ combination matrix. 59 suitable matches were made above threshold, with position and time resolutions of 14 mm and 80 ps. The efficiency of our disambiguation technique on this LAPPD was found to be 59 pairs out of 141 PEs, or 42\%.

\subsection{ANNIE - Water Cherenkov Detector}

\begin{figure}[!htbp]
\includegraphics[width=1\linewidth]{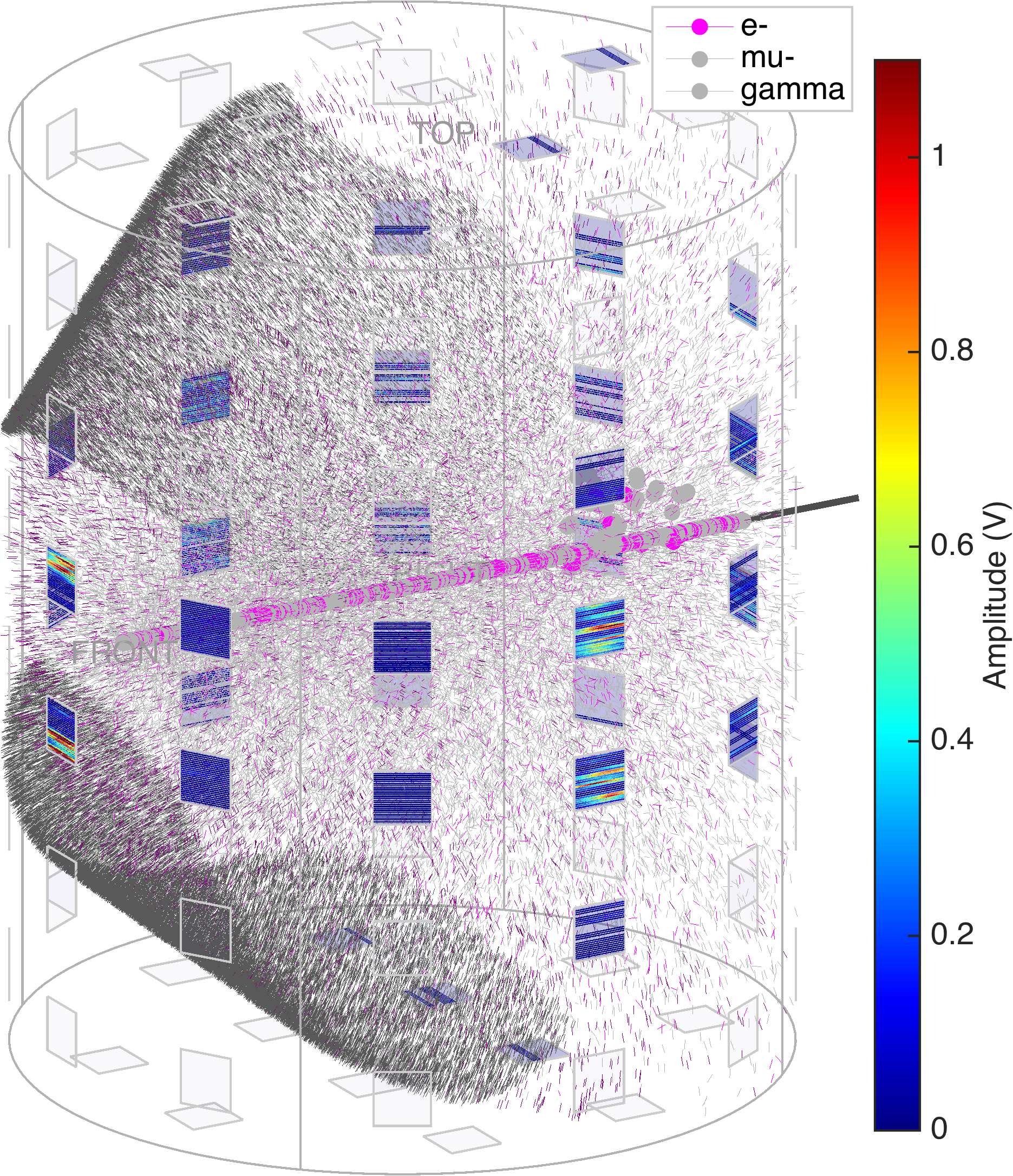}
\centering
\caption{Simulated event display for a 10 GeV muon traversing the ANNIE\cite{ANNIE} water Cherenkov detector. The image is captured 12 ns after the muon enters the detector. About 2500 PEs are collected over the 96 ANNIE LAPPDs per muon event.}
\label{ANNIE}
\end{figure}

\begin{figure}[!htbp]\centering
\includegraphics[width=1\linewidth]{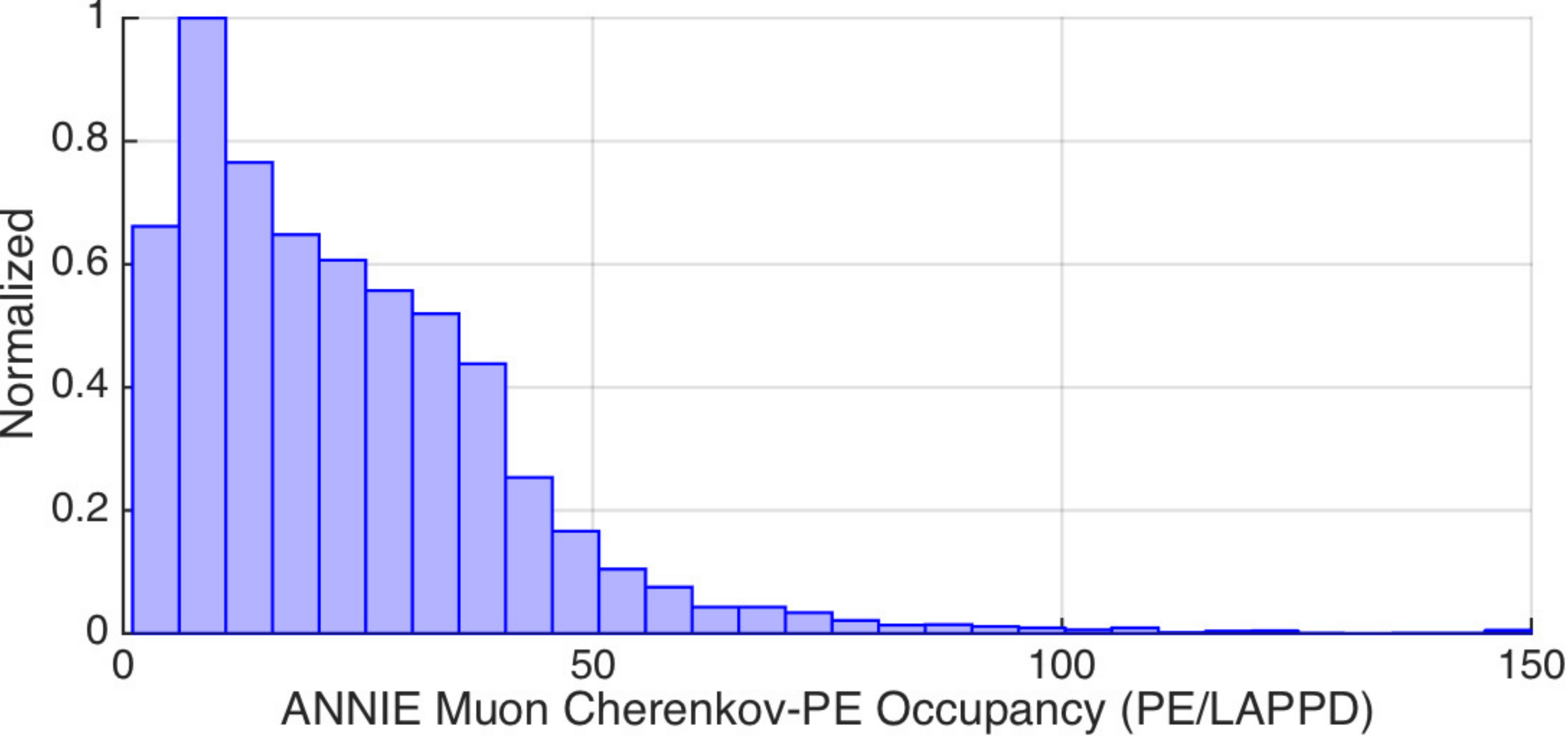}
\caption{GEANT4\cite{GEANT} and MATLAB\cite{MATLAB} MonteCarlo results of LAPPD PE occupancy for muons traversing ANNIE, showing that most LAPPDs receive about 10-20 PEs, though a small fraction are saturated by $>75$.}
\label{ANNIE_occupancy}
\end{figure}

\begin{figure}[!htbp]\centering
\includegraphics[width=1\linewidth]{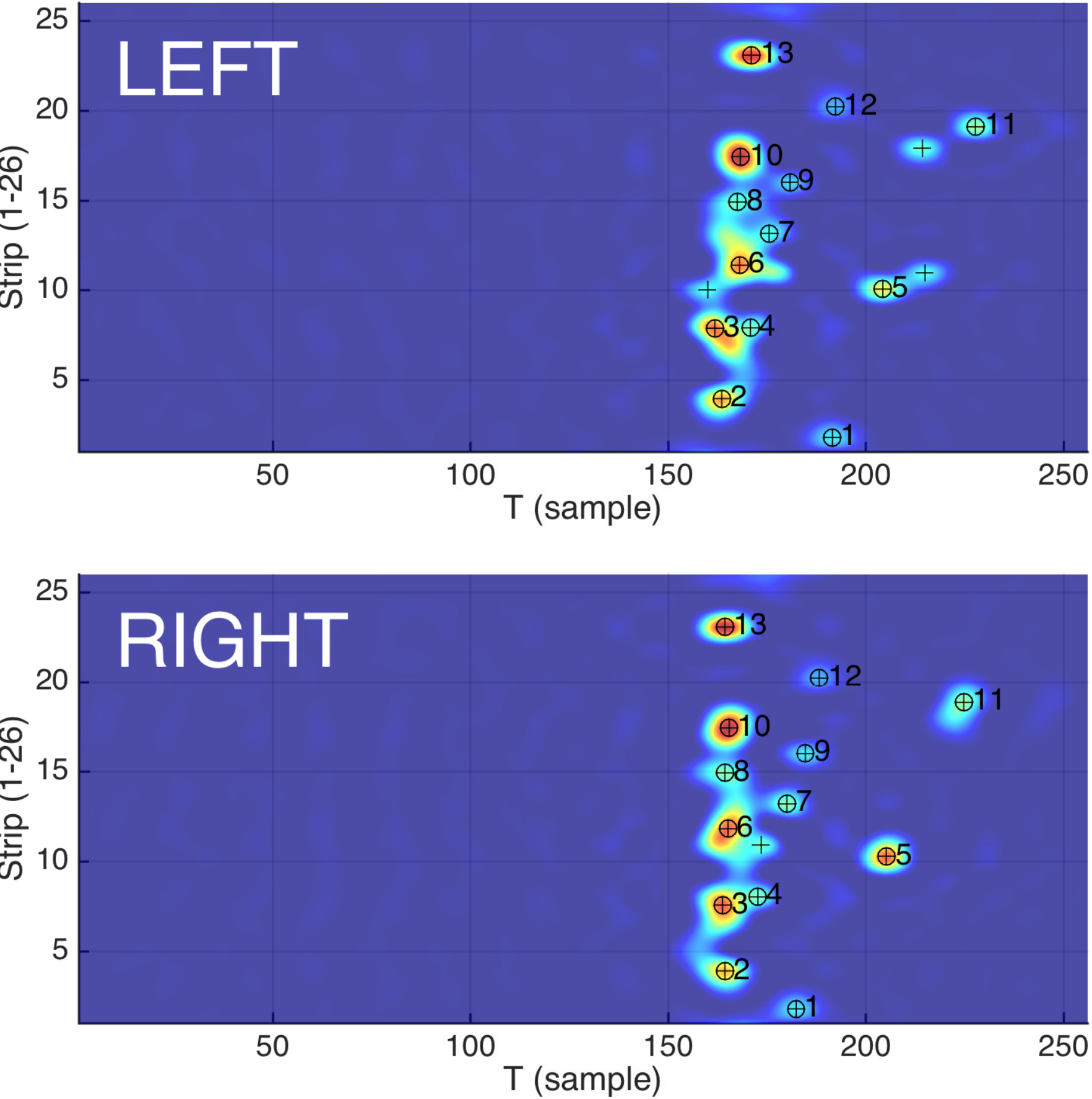}
\caption{ANNIE water-Cherenkov disambiguation example showing muon PE depositions on 1 single LAPPD (LEFT and RIGHT anode strip sides shown). Crosses indicate resolved PEs, and circles indicate successful left-right pairing of resolved PEs. 13 matches were found from the 16$\times$14 pulse matrix $\Lambda_{ij}$. Match position and time resolutions are 12 mm and 65 ps at 30\% efficiency (13 matches / 43 PEs).}
\label{ANNIE_Disambiguation}
\end{figure}

The Accelerator Neutrino Neutron Interaction Experiment\cite{ANNIE} (ANNIE), shown in Figure \ref{ANNIE}, is a large water-Cherenkov detector being deployed at the Fermilab beam line to search for neutron multiplicities in high energy interactions. Up to 100 LAPPDs deployed around the edge of the detector will measure Cherenkov light. We use our model to determine PE densities (shown in Figure \ref{ANNIE_occupancy}) and disambiguation capabilities (Figure \ref{ANNIE_Disambiguation}) of the proposed design.

\section{Disambiguation Trade Study: 1 to 100 PEs}

\subsection{Efficiency and Resolution Metrics}
The information `linking' a true PE to it's causal original `true' PE is lost during our disambiguation process, and is unrecoverable at the higher occupancies as many times a `reconstructed' PE is actually made up of several `true PEs'. The alternative we use instead is free matching from each reconstructed PE to the nearest true PE. `Nearest' in this case means the shortest 4D $[dx dy dz dt]$ distance, where $dx$ is in mm and $dt$ is converted from ns to mm at a rate of 180 mm/ns ($c$ / scintillator refractive index of $n\approx1.6$). We do not enforce unique matches. Every reconstructed PE finds a match, but not every true PE is matched as the true PE count nearly always exceeds the reconstructed PE count (very few false positives).

Position resolution is measured as the $1\sigma$ of the euclidean 2D ranges between the true and reconstructed PEs on the surface of the LAPPD. Time resolution is measured as the $1\sigma$ of the $dt$ between true and reconstructed PE pairs. Efficiency is simply the ratio to reconstructed to true PE counts. If a pulse is larger than twice the mean amplitude, it is broken down into a number of smaller pulses consistent with the total amplitude observed, each with identical position and time estimates. 1000 MCs are run for each table row.

\subsection{Nominal LAPPD Anode}
To see how PE resolutions and efficiencies are impacted by increased PE density on the LAPPD we run a Trade Study (TS) over PE count. We collect PE's on one of the scintillation detector LAPPDs from a centered, isotropic scintillating point source. Photon emission times are consistent with Eljen EJ-254 plastic scintillator (0.9 ns rise, 2.2 ns fall time). The point source is 7.5 cm distant from the LAPPD center.

We run 1000 Monte Carlo (MC) simulations at each TS point to determine the PE position $x$ and time $t$ resolutions, as well as the disambiguation efficiency. This efficiency is simply the number of matched left-right pulse pairs divided by the total number of PEs on the LAPPD. For example if 5 PEs land on the LAPPD and we resolve 3 matches, our disambiguation efficiency would be 0.60.

\begin{table} [!htbp]
\fontsize{10}{10}\selectfont
\begin{center}
\begin{tabular}{l|l|l|l}
\bf PE $\#$ 	  	&  efficiency 	& \bf $x$ {\bf(mm)} & \bf $t$ {\bf(ns)} \\
\hline \bf      1 	&   1.00		&    0.9 			&  0.060 \\
\hline \bf      5 	&   0.95		&    2.2 			&  0.064 \\
\hline \bf     25 	&   0.74		&    5.1 			&  0.079 \\
\hline \bf    100 	&   0.53		&   12.8 			&  0.108 \\
\end{tabular}
\caption{{\bf 26-strip LAPPD} PE position $x$ and time $t$ MC $1\sigma$ resolutions and efficiencies for a scintillating point source in the center of the neutron detector. 60 ps  TTS\cite{LAPPDtiming} included in these numbers. Position resolution $x$ combines the 2 along and across-strip resolutions into a single range resolution.}
\label{table:1}
\end{center}
\end{table}

Table \ref{table:1} shows the result of our LAPPD scintillation PE count trade study. Measurement resolutions worsen and efficiencies drop as PE count increases. This drop in performance is probably due to mismatches in PE pairing in the $\Lambda_{ij}$ matrix, as well as inability to deconvolve heavily overlapping PEs in signal space $S$. Overall  spatial and temporal resolutions seem to remain acceptable for many scintillator detector applications such as particle vertex tracking.

\subsection{Double-Density LAPPD Anode}

Table \ref{table:2} shows the performance of a hypothetical 52-strip LAPPD variant in comparison to the 26-strip LAPPD in Table \ref{table:1}. This 52-strip variant has twice the strip count, half the strip width, and half the charge cloud size (2 mm vs 4 mm). The charge cloud size may be manipulated by varying the gaps between the MCP layers and the anode strips. A shorter gap should reduce the charge cloud size, which should maximize the benefit of an increased strip count for the purposes of disambiguating higher occupancy events.

As expected, we find these modifications produce a dramatic improvement in performance when comparing Table \ref{table:1} and Table \ref{table:2}, particularly in spatial resolution.

\begin{table} [!htbp]
\fontsize{10}{10}\selectfont
\begin{center}
\begin{tabular}{l|l|l|l}
\bf PE $\#$ 		&  efficiency 	& \bf $x$ {\bf(mm)} & \bf $t$ {\bf(ns)} \\
\hline \bf      1 	&   0.98		&    0.6 			&  0.060 \\
\hline \bf      5 	&   0.98		&    1.3 			&  0.062 \\
\hline \bf     25 	&   0.84		&    3.9 			&  0.075 \\
\hline \bf    100 	&   0.64		&    7.8 			&  0.087 \\
\end{tabular}
\caption{{\bf 52-strip LAPPD} PE position $x$ and time $t$ MC $1\sigma$ resolutions and efficiencies for a scintillating point source in the center of the neutron detector.  60 ps TTS\cite{LAPPDtiming} included in these numbers. Position resolution $x$ combines the 2 along and across-strip resolutions into a single range resolution.}
\label{table:2}
\end{center}
\end{table}

\section{Conclusion}

In this paper we have established a method for solving multiple PE arrivals simultaneously on stripline-anode MCPs. Our disambiguation method relies on comparisons of left-right pulse amplitude similarity $\Lambda_a$, time similarity $P_t$, and location similarity $\Lambda_y$.

We show modeled performance metrics for small scintillator detectors as well as large water-Cherenkov detectors. We note increased difficulties with Cherenkov disambiguation (Figs. \ref{ANNIE}, \ref{ANNIE_Disambiguation}, \ref{a1}, \ref{a2}, \ref{a3}) vs. scintillation PE disambiguation (Figs. \ref{NTC}, \ref{deconv3}, \ref{a4})  due to the tighter arrival time distributions of the Cherenkov light. Our trade study results show reliable disambiguation of scintillation PEs at levels of up to 4 PEs/strip, and LAPPD time and position resolutions which should suffice for many real-world particle physics applications. We conclude that our disambiguation method is a reliable, practical way to expand the performance envelope of stripline anode MCPs such as the LAPPD into higher PE density territory than the single-PE occupancies they were initially conceived for.

\section{Acknowledgements}

We would like to thank Henry Frisch at the University of Chicago for his guidance on LAPPD-related matters, as well as Eric Oberla and Andrey Elagin at the University of Chicago for their expertise during the LAPPD laser collections at Argonne National Lab in 2014. This material is based upon work supported by the National Geospatial-Intelligence Agency and the University of Chicago.

\appendix
\section{}

\begin{figure}[!htbp]\centering
\includegraphics[width=1\linewidth]{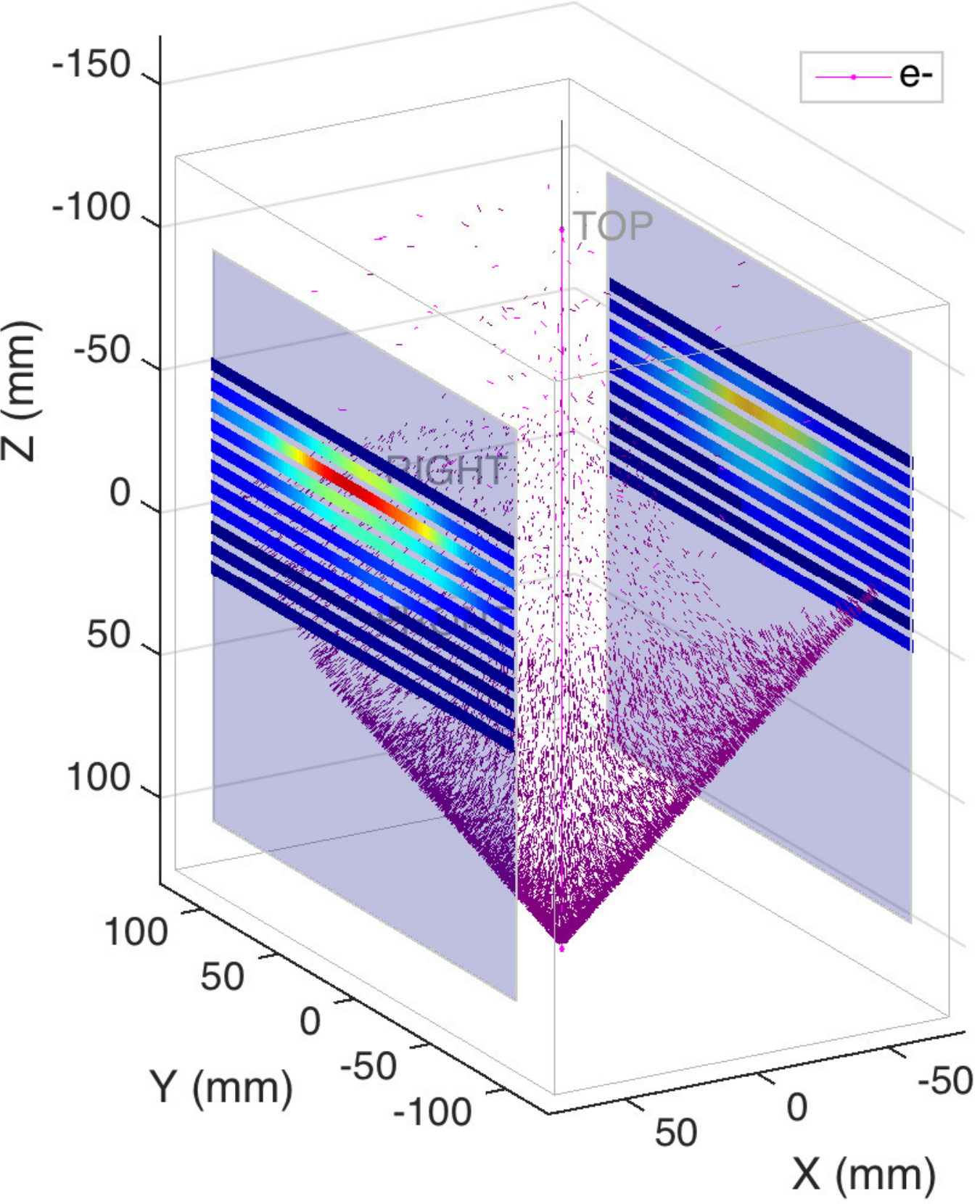}
\caption{Example Cherenkov event with a line source passing straight through the center of a water detector with LAPPDs on either side.}
\label{a1}
\end{figure}

\begin{figure*}[!htbp]\centering
\includegraphics[width=1\linewidth]{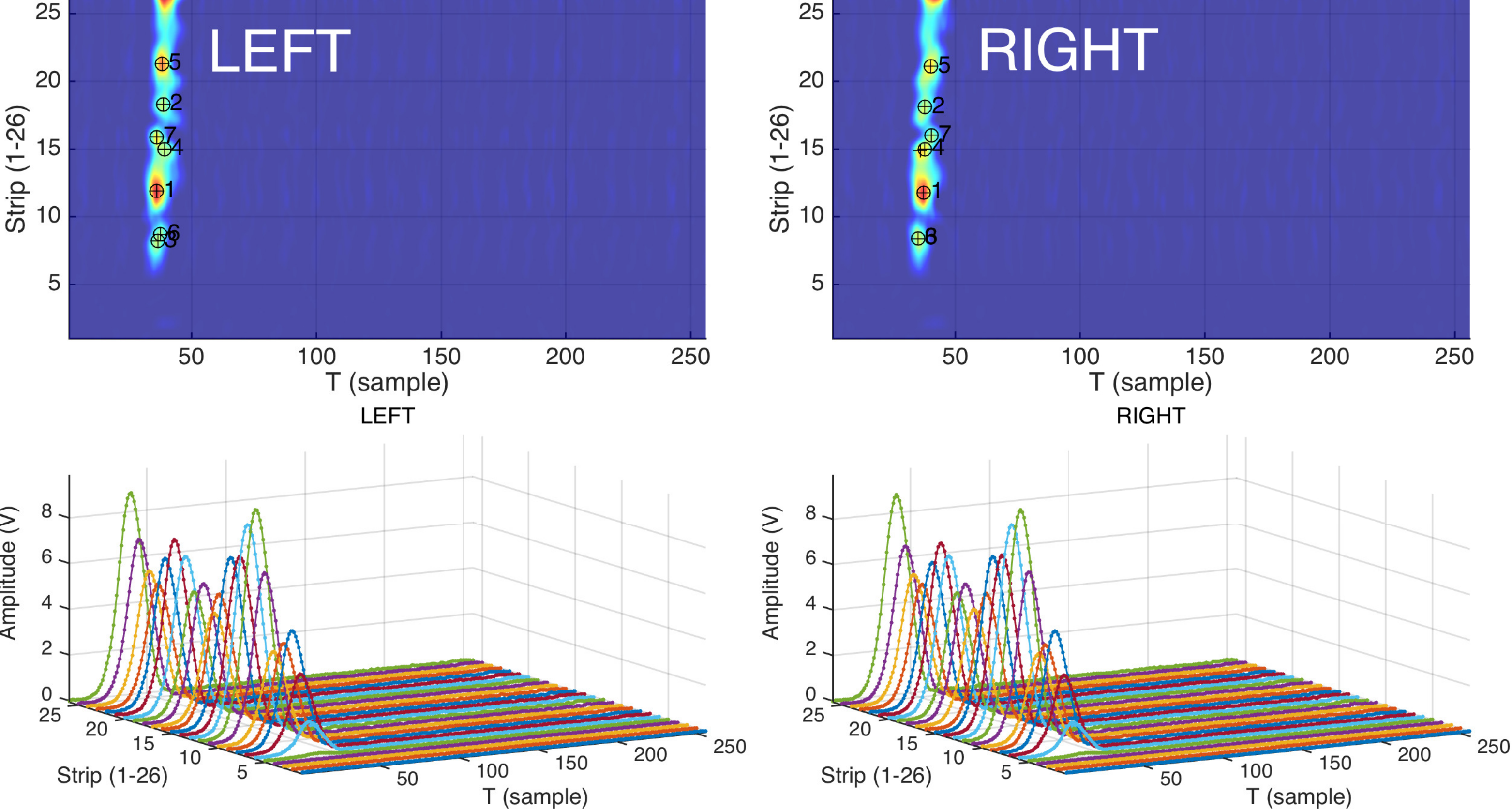}
\caption{Example 100 PE Cherenkov event as seen on the 26 LAPPD signal strips. {\bf Clipping is turned off here}, otherwise the signals would hit the PSEC4 dynamic range ceiling of about 1.1 V. If read out by a PSEC4 chip, nearly every strip would saturate here. Even though 100 PEs created the signals seen on the 26 strips, only about 10 peaks are resolvable in the spline matrix unfortunately. This is largely due to the tight distribution in time of the arriving Cherenkov light. Crosses indicate resolved PEs, and circles indicate successful left-right pairing of resolved PEs.}
\label{a2}
\end{figure*}

\begin{figure*}[!htbp]\centering
\includegraphics[width=1\linewidth]{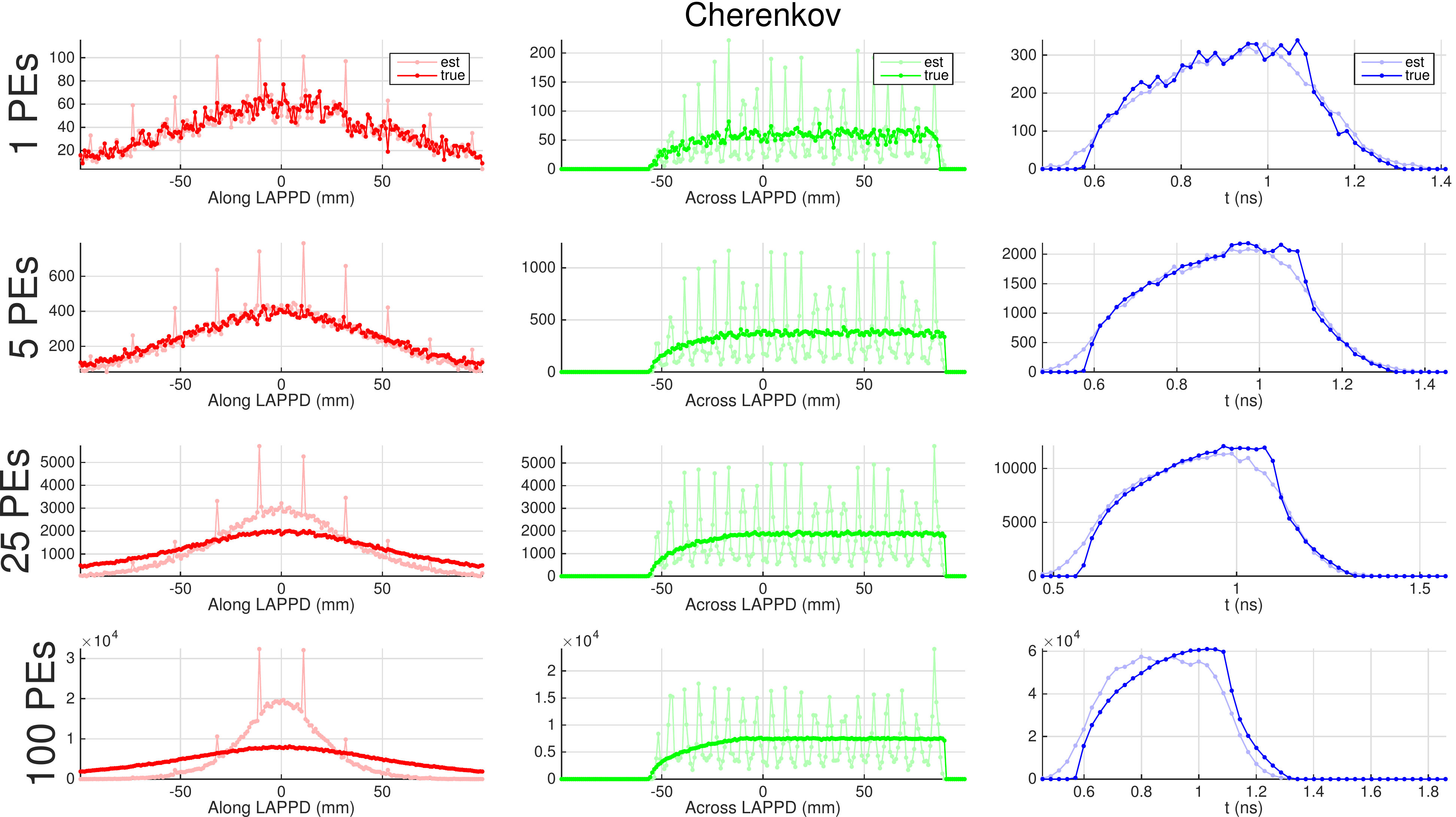}
\caption{Distribution of true and reconstructed PEs shown for 1, 5, 25 and 100 PE {\bf Cherenkov} events as seen on the 26 LAPPD signal strips. {\bf Clipping is turned off here.} Timing smeared by 60 ps TTS.}
\label{a3}
\end{figure*}

\begin{figure*}[!htbp]\centering
\includegraphics[width=1\linewidth]{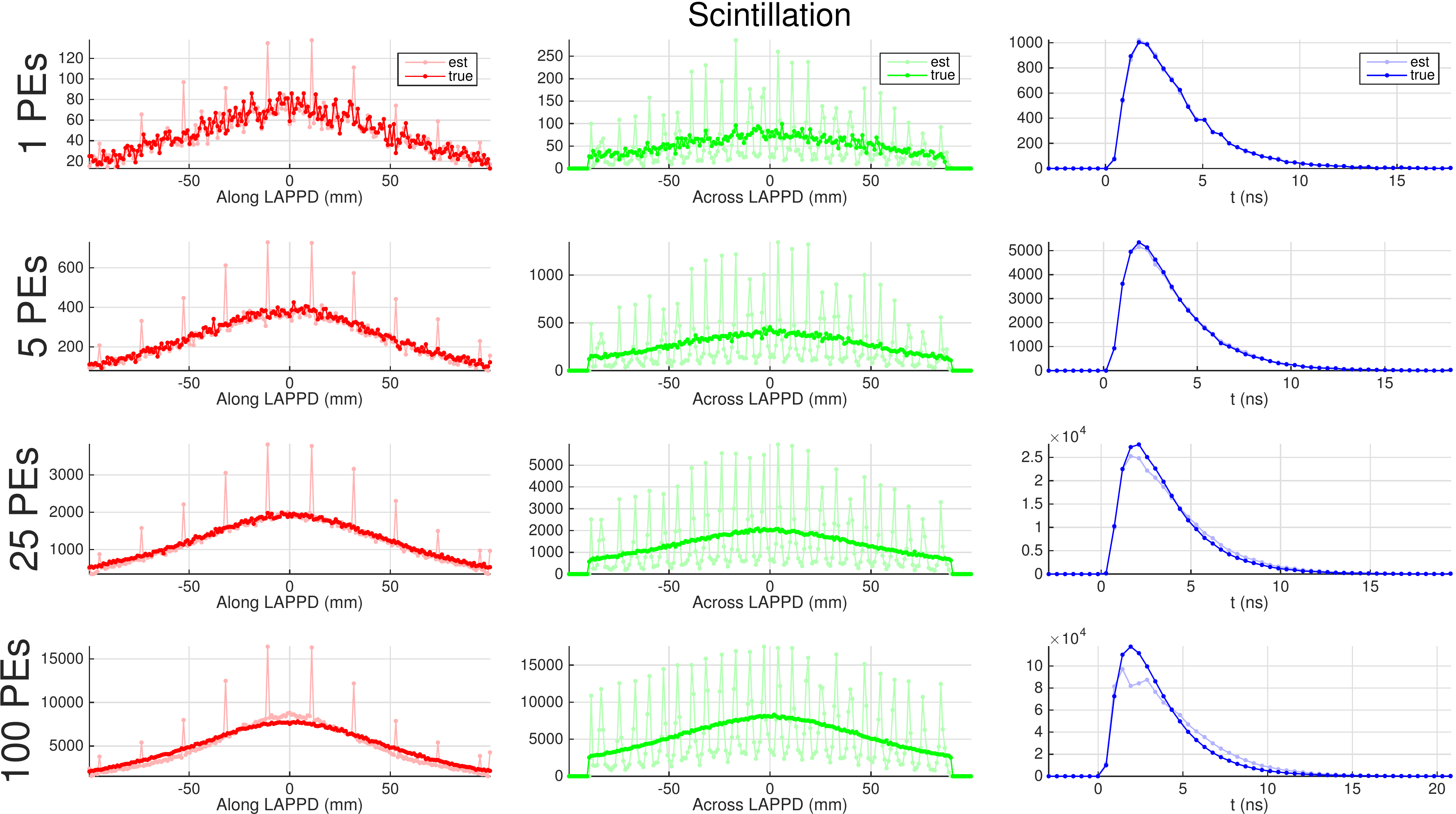}
\caption{Distribution of true and reconstructed PEs shown for 1, 5, 25 and 100 PE {\bf Scintillation} events as seen on the 26 LAPPD signal strips. {\bf Clipping is turned off here.} Timing smeared by 60 ps TTS.}
\label{a4}
\end{figure*}


\section*{References}
\begin{thebibliography}{00}

\bibitem{Lampton_1987} Lampton, M. and Siegmund, O. and Raffanti, R., ``Delay line anodes for microchannel‐plate spectrometers" Review of Scientific Instruments, 58, 2298-2305 (1987), \url{http://dx.doi.org/10.1063/1.1139341}

\bibitem{Jagutzki_2007} Ottmar Jagutzki, Volker Dangendorf, Ronald Lauck, Achim Czasch and James Milnes ``A position- and time-sensitive photon-counting detector with delay-line readout", Proc. SPIE 6585, Optical Sensing Technology and Applications, 65851C (2007), \url{http://dx.doi.org/10.1117/12.721933}

\bibitem{Rindi_1970} Rindi, A. and Perez-Mendez, V. and Wallace, R. I., ``Delay line readout for proportional chambers" Nuclear Instruments and Methods, 77, 325-327 (1970)

\bibitem{Grove_1970} Grove, R. and Lee, K. and Perez-Mendez, V. and Sperinde, J., ``Electromagnetic delay line readout for proportional wire chambers" Nuclear Instruments and Methods, 77, 325-327 (1970)

\bibitem{anodepaper}  Grabas, H. and Obaid, R. and  Oberla, E. and Frisch, H. and Genat, J.-F. and Northrop, R. and  Tang, F. and McGinnis, D. and Adams, B. and Wetstein, M., ``RF strip-line anodes for psec large-area mcp-based photodetectors" Nuclear Instruments and Methods, A 711 (0) (2013) 124-131  \url{doi:http://dx.doi.org/10.1016/j.nima. 2013.01.055}

\bibitem{GEANT} GEANT4 - A Simulation Toolkit, S. Agostinelli \textit{et al.}, Nuclear Instruments and Methods A 506 (2003) 250-303 \url{http://geant4.cern.ch/}

\bibitem{MATLAB} MathWorks, MATLAB R2015a, computer program, The MathWorks Inc., Natick, MA, USA. \url{http://www.mathworks.com}

\bibitem{EJ-254} Eljen Technology. EJ-254 Boron Loaded Plastic Scintillator,  \url{http://www.eljentechnology.com/}

\bibitem{test setup} B.W. Adams, M. Chollet, A. Elagin, R. Obaid, E. Oberla, A. Vostrikov, M. Wetstein, and P. Webster. A test-facility for large-area microchannel plate detector assemblies using a pulsed sub-picosecond laser. Rev. Sci. Instrum., 84, 2013, \url{http://dx.doi.org/10.1063/1.4810018}

\bibitem{PSEC4 Paper} E. Oberla, H. Grabas, J.-F. Genat, H.J. Frisch, K. Nishimura, and G. Varner. A 15 GSa/s, 1.5 GHz bandwidth waveform digitizing ASIC. Nucl. Instr. Meth A, 735:452–461, 2014, \url{http://dx.doi.org/10.1016/j.nima.2013.09.042}

\bibitem{ANNIE} Letter of Intent: The Accelerator Neutrino Neutron Interaction Experiment (ANNIE), April 2015, \url{http://arxiv.org/abs/1504.01480}

\bibitem{WNR} ``Digital Image Processing", R. C. Gonzalez and R. E. Woods, Addison-Wesley Publishing Company, Inc., 1992.

\bibitem{LAPPDtiming} B.W. Adams, A. Elagin, H.J. Frisch, R. Obaid, E. Oberla, A. Vostrikov, R.G. Wagner, J. Wang, M. Wetstein, Timing characteristics of Large Area Picosecond Photodetectors, Nuclear Instruments and Methods in Physics Research Section A, Volume 795, 21 September 2015, Pages 1-11, ISSN 0168-9002, \url{http://dx.doi.org/10.1016/j.nima.2015.05.027}

\end{thebibliography}
\end{document}